\documentclass{nature}
\usepackage[utf8]{inputenc}
\usepackage[T1]{fontenc}
\usepackage{graphicx}
\usepackage{xcolor}
\usepackage{physics}
\usepackage{bm}
\usepackage{amsfonts}
\usepackage{soul}
\usepackage{hyperref}
\hypersetup{
    colorlinks=true,
    linkcolor=blue,
    filecolor=blue,
    urlcolor=blue,
   citecolor=blue,
}

\usepackage{upgreek}
\makeatletter
\let\saved@includegraphics\includegraphics
\AtBeginDocument{\let\includegraphics\saved@includegraphics}
\renewenvironment*{figure}{\@float{figure}}{\end@float}
\makeatother
\title{Room temperature exciton-polariton neural network with perovskite crystal}
\author{Andrzej Opala$^{1,2,\dagger}$, Krzysztof Tyszka$^{1,{\dagger}}$, Mateusz Kędziora$^{1}$, Magdalena Furman$^{1}$, Amir Rahmani$^{2}$, Stanisław Świerczewski$^{1}$, Marek Ekielski$^{3}$, Anna Szerling$^{3}$, Michał Matuszewski$^{2,4}$ and Barbara Piętka$^{1,*}$}
\begin{document}
\begin{spacing}{1.125}
\maketitle
\begin{scriptsize}
\begin{affiliations}
\item Institute of Experimental Physics, Faculty of Physics, University of Warsaw, ul. Pasteura 5, PL-02-093 Warsaw, Poland
\item Institute of Physics, Polish Academy of Sciences, Aleja Lotników 32/46, PL-02-668 Warsaw, Poland
\item Łukasiewicz Research Network - Institute of Microelectronics and Photonics, al. Lotników 32/46, 02-668, Warsaw, Poland
\item Center for Theoretical Physics, Polish Academy of Sciences Aleja Lotników 32/46, 02-668 Warsaw, Poland
\\
${\dagger }$- equally contributed
\\
${*}$- corresponding author: barbara.pietka@fuw.edu.pl
\end{affiliations}

\end{scriptsize}

\begin{abstract}
Limitations of electronics have stimulated the search for novel unconventional computing platforms that enable energy-efficient and ultra-fast information processing. Among various systems, exciton-polaritons stand out as promising candidates for the realization of optical neuromorphic devices. This is due to their unique hybrid light-matter properties, resulting in strong optical nonlinearity and excellent transport capabilities. However, previous implementations of polariton neural networks have been restricted to cryogenic temperatures, limiting their practical applications. In this work, using non-equillibrium Bose-Einstein condensation in a monocrystalline perovskite waveguide, we demonstrate the first room-temperature exciton-polariton neural network. Its performance 
is verified in various machine learning tasks, including binary classification, and object detection.
Our result is a crucial milestone in the development of practical applications of polariton neural networks and provides new perspectives for optical computing accelerators based on perovskites.
\end{abstract}
\vspace{1mm}

Neuromorphic computing systems (NCSs) are hardware-based computing devices inspired by the structure and function of biological neurons. Unlike traditional software-based ANNs, where neurons are simulated by algorithms~\cite{Markovic_2020}, neurons in NCSs are implemented directly in hardware. This allows for the simultaneous storage and processing of information, resulting in computational parallelism and overcoming the limitations of Turing-type electronics. In particular, photonic NCSs use optical elements to process information encoded in the degrees of freedom of light, including intensity, polarization, and phase~\cite{Zuo_2022}. Over the last few decades, realizations of neuromorphic devices based on different optical platforms, such as nanophotonic circuits~\cite{Shen_2017}, phase-change materials~\cite{Feldmann_2019}, and disordered media~\cite{Wang_2024}, among others~\cite{brunner_18}, have made significant progress~\cite{Li_2024}. Despite these advances, optical neuromorphic computing still faces challenges. A significant obstacle in creating nonlinear hardware neurons is the lack of photon-photon interaction. To address this problem, many optical hardware neural networks have adopted a hybrid optoelectronic approach, where optical elements handle information transfer while electronic components perform nonlinear processing. This hybrid solution combines the speed of optical processing with the flexibility of electronic systems, including tunable nonlinearity. However, this is a remedy only for one of the issues, since optoelectronic data conversion is itself a bottleneck limiting the ability to provide energy-efficient computing~\cite{Matuszewski_2024}. A promising alternative for the hybrid approach involves the use of exciton-polaritons (hereafter polaritons), which are half-light, half-matter bosonic quasiparticles providing stronger light-matter interactions~\cite{Opala_23, Kavokin_2022}.

Polaritons combine electronic and photonic properties in a single quasiparticle, forming a unified, optically controlled physical platform~\cite{Deng_2010, Byrnes_2014}. They are formed when the coherent exchange of energy between an elementary electronic excitation of a semiconductor crystal (exciton) and a quantum of the electromagnetic field (photon) occurs on a time scale significantly shorter than its lifetime, which is observed in the so-called strong coupling regime~\cite{Hopfield_1958, Weisbuch_1992}. In these conditions, the two constituents lose their distinct nature and form a bosonic quasiparticle remaining in a coherent superposition. Polaritons exhibit bosonic statistics and are capable of undergoing a phase transition to a non-equilibrium bosonic condensate~\cite{Kasprzak_2006} even at room temperature~\cite{Sanvitto_2016}. The photonic component of polariton results in an ultra-low effective mass, leading to excellent transport properties, when the excitonic component of polaritons gives rise to significant interparticle interactions. From the perspective of applications, polaritons have been used for creating optical lattices~\cite{Alyatkin_2021}, simulators~\cite{Berloff_2017, Tao_2022}, energy-efficient, ultrafast optical switches~\cite{Fraser_2017, Genco_2024}, logic gates~\cite{Mirek_2021,Byrnes_2024, Sannikov_2024}, routers~\cite{Bloch_2015}, transistors~\cite{Ballarini_2013, Zasedatelev_2019}, and neural networks~\cite{Opala_23, Kavokin_2022}. Nevertheless, neural network implementations have been limited to cryogenic temperatures. 
Therefore, achieving a polariton neural network that operates at room temperature remained one of the major challenges in the field of optical computing with exciton-polaritons.

This work overcomes this challenge by using lead halide perovskites, an emerging semiconductor material, to realize the first polariton neural network operating at room temperature. In our implementation, we use separated polariton neurons whose activation is correlated with the transition to the non-equilibrium Bose-Einstein condensate. We demonstrate the performance of the proposed neuromorphic system in machine learning tasks such as object detection and binary classification. In the first task, we employ polariton NCS to distinguish images of different geometrical objects. The analyzed objects fell into four categories: small circles, large circles, squares, and empty planes. This task is particularly challenging for any linear classifier due to the unspecified object positions. However, based on experimental results and considering the real noise, we show that a polariton neural network with only four neurons, whose activation functions resemble a parametric rectified linear unit (ReLU), is able to solve this problem. We classified 4,000 objects with an accuracy of $\sim$96$\%$ (experimental inference), higher than the average accuracy obtained with an ideal linear classifier, more than $\sim$25$\%$ and lower than software prediction by only $\sim$3$\%$. These results underscore the impact of polariton high nonlinearity, induced by the possibility of condensation at room temperature, and the relevance of polariton neuromorphic systems in machine learning applications. Next, we validated the perovskite-based polariton neural network on a binary classification problem with linearly inseparable datasets.  We achieved classification accuracies that outperformed the benchmark linear classifier accuracy by more than 92$\%$ for three different types of datasets with varying levels of complexity. 

\subsection{Room temperature exciton-polariton neural network.} 
Polariton neurons are characterized by a strong nonlinearity, resulting from polariton condensation and strong interparticle interactions mediated by their excitonic component. Previous experimental implementations of polariton neural networks, inspired by the theoretical concepts~\cite{Ortega_2015, Byrnes_2013,opala2019neuromorphic}, have used optical signals modulated in space, time, or both, to activate polariton neurons. These systems have been successfully developed in both recurrent and feed-forward neural network architectures and have been tested on tasks such as handwritten digit and spoken word classification, demonstrating high energy efficiency~\cite{Ballarini_2020,Mirek_2021,Mirek_2022,Opala_2022}. The mentioned realizations of polariton NCS have been limited to optical microcavities based on gallium arsenide and cadmium telluride semiconductors, which exhibit strong nonlinear effects. However, because of the low exciton binding energies and low exciton-photon coupling strength, the applicability of these materials is typically limited to cryogenic temperatures.
Finding a solution to this problem by harnessing emerging semiconductor materials was pointed out as a crucial step to achieving real-world applications of polariton neuromorphic system~\cite{matuszewski2021energy, Kavokin_2022}. 

\begin{figure}
     \centering
        \centering  \includegraphics[width=1\textwidth]{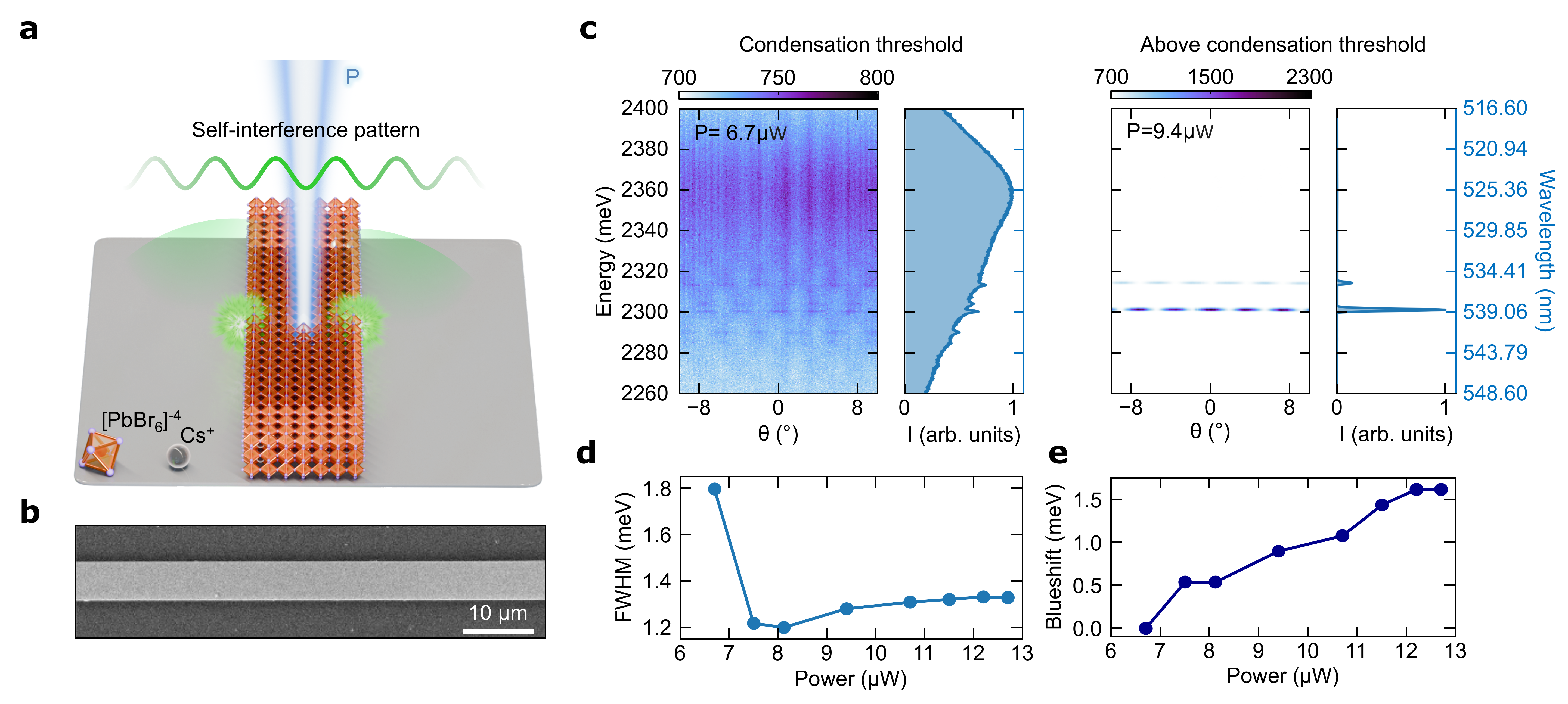}
    \caption{{\bf{Room temperature exciton-polariton condensation in perovskite waveguide.}} {\bf{a}} Scheme of polariton condensation within the perovskite waveguide, where green dots mark the emission points at the waveguide edges, resulting in a self-interference pattern during emission detection. {\bf{b}}, Scanning electron microscopy (SEM) image of the perovskite waveguide. {\bf{c}}, Angle-resolved spectra of the perovskite emission, showing measurements below and above the condensation threshold (left panels) presented together with the normalized emission spectrum averaged over the measured angles (right panels). {\bf{d}}, Power-dependent full width at half maximum (FWHM) and blueshift {\bf{e}}, of the most intense condensation mode, confirming the appearance of the polariton condensation process.}
    \label{fig_0}
\end{figure}

Among all materials capable of polariton condensation at room temperature, perovskites based on lead halides (henceforth perovskites) stand out for their robustness, easy synthesis, and scalability. The nonlinear properties of perovskite polaritons are used in the fabrication of lasers~\cite{Zhang2014}, Hamiltonian simulators~\cite{Tao_NatMat2022} and are being considered for integrated photonic circuits~\cite{Karabchevsky2023}. Perovskites are also distinguished by their ability to achieve strong coupling without troublesome encapsulation into microcavity, particularly in the case of high-quality microcrystals with well-defined edges and planar shapes, such as wires or plates. Several methods have been developed for obtaining such crystals in both nanometer~\cite{Zhu2015} and micrometer sizes~\cite{Zhang2014}. In addition to randomly distributed perovskite crystals, it is also possible to perform scalable synthesis of crystals in the form of microplates~\cite{Feng2016}, microwires~\cite{Li_2024}, or any predefined shape~\cite{Kedziora2024}. This possibility paves the way for the fabrication of reproducible and scalable polariton platforms for neuromorphic computing at room temperature. To demonstrate this, we use polariton condensate in CsPbBr$_3$ perovskite waveguide (see Figure~\ref{fig_0}a-b), as the physical platform to implement an analogue polariton neural network. A notable benefit of this approach is eliminating the necessity for additional Bragg mirrors, which reduces both the financial cost and the technical difficulty associated with the manufacturing process.
The perovskite microwires employed in the experiment were synthesized using the template-assisted crystallization method. 

The fabricated microwires exhibit characteristic edge emission when excited at the center of the wire by a non-resonant laser, as schematically shown in Fig.~\ref{fig_0}a. The perfect crystal facets at the wire edges create a Fabry-Pérot resonator, supporting standing cavity modes across the wire and waveguide modes along its length. Due to the strong excitonic resonance in perovskites, these modes are strongly coupled with the excitonic resonance, forming waveguided polariton modes~\cite{Kedziora2024}.  Fig.~\ref{fig_0}c shows the emission spectra captured in the far field, both below and above the condensation threshold. The spectrally broad emission below the condensation threshold, centered at 2360 meV (530~nm), corresponds to excitonic emission in the perovskite. Above the threshold, the emission spectrum consists of two peaks, with the lowest energy peak becoming dominant as excitation power increases. This originates from the formation of a polariton condensate at waveguided modes, resulting in non-linear light emission from the crystal edges. The angle-resolved interference pattern visible in Fig.~\ref{fig_0}c confirms the formation of a coherent state across the perovskite waveguide. Panels d and e display quantitative trends in the emission peak blueshift and full width at half maximum (FWHM) as excitation power increases, highlighting the clear onset of condensation and providing a distinct fingerprint of the interactions within the system~\cite{Kasprzak_2006}.

\begin{figure}
     \centering
    \centering  \includegraphics[width=0.75\textwidth]{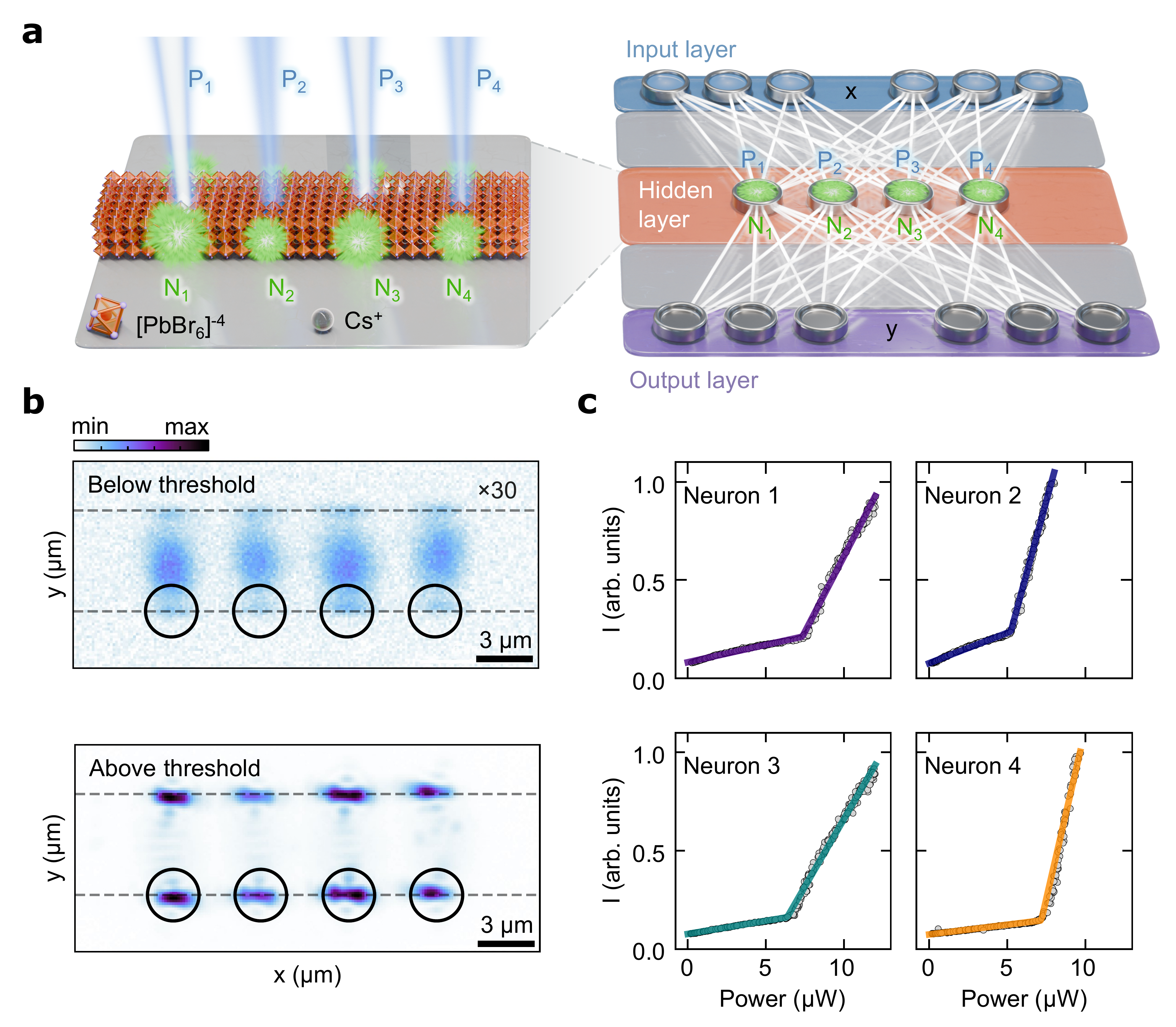}
    \caption{\textbf{Feed-forward neural network based on a perovskite waveguide.} Panel~\textbf{a} shows a scheme of a feed-forward neural network containing four polariton neurons in the hidden layer, labelled as $N_i$, realized in a single polariton waveguide. The waveguide is excited by laser pulses sequentially directed at spatially separated positions along the microwire, with an intensity $P_i$. Panel~\textbf{b} depicts the integrated polariton emission both below and above the condensation threshold, for the implementation of the experiment shown in panel~\textbf{a}, where four laser pulses are applied along a single perovskite wire marked with dashed lines. Panel~\textbf{c} shows the measured nonlinear photoluminescence intensity of four polariton nodes. The measured characteristics resemble the shapes of parametric ReLU functions, which are shown as thick solid lines. 
}
    \label{fig_1}
\end{figure}

Figure~\ref{fig_1}a presents the concept of an experiment where a single-crystal perovskite microwire acts as a hidden layer of a neural network. In this configuration, four ($i=1...4$) spatially and temporally separated laser pulses, with intensities denoted as $P_i$, are used to create analog polariton neurons, i.e., condensation sites, through excitation with non-resonant laser pulses. Optical control of the neurons is achieved by adjusting the excitation power.

To demonstrate the feasibility of this proposed implementation, a single microwire was excited with four excitation beams along the perovskite waveguide. The photoluminescence maps both below and above the condensation threshold are illustrated in Fig.~\ref{fig_1}b. The small time delay between each neuron excitation ensured mutual isolation, preventing interference and feedback effects between condensation sites within this hidden layer. The details of the experimental scheme can be found in Fig.~S2 in the SI. 

The implementation of the experiment (setup shown in Fig.~S1 in SI), which was realized in training and inference phases, involved measuring emission from four distinct positions on the perovskite microwire, each time focusing on a single position along the microwire. Example of a nonlinear input-output characteristics of each neuron is illustrated in Fig.~\ref{fig_1}c. The measurements were carried out in sequences in which the optical power of the excitation was repeatedly linearly varied between 0~$\upmu$W and 14~$\upmu$W. A single scan consisted of 122 samples. 

At all four measured positions, the neural activation threshold, equivalent to polariton condensation threshold, occurred between 5.25~$\upmu$W and 7.35~$\upmu$W. Different condensation threshold ($P_{\rm th}$) and the intensity slope for each neuron could be attributed to slight variations in sample or crystal properties, such as differences in the local structure quality or slight variations in the position of the laser beam relative to the longitudinal axis of the wire. Each of the nonlinear optical polariton responses was fitted to the parametric ReLU function, given by 
\begin{equation} \label{eq:phi}
    \varphi_i(P_i) = 
\begin{cases} 
\alpha_i (P_i - P^{\rm th}_i) + b_i & \text{if } P_i < P^{\rm th}_i \\
\beta_i (P_i - P^{\rm th}_i) + b_i & \text{if } P_i > P^{\rm th}_i
\end{cases},
\end{equation}
where, $\alpha_i$, $\beta_i$, $b_i$, and $P_i^{\text{\rm th}}$ are parameters obtained by fitting the experimental data. For a detailed explanation of the above equation, please see the Method section. 

During the fitting process, the initial six scans of the optical polariton responses, are employed to fit four activation functions. The fits are illustrated in Fig.~\ref{fig_1}c on top of the experimental data presented in grey dots. 
The stability of optical activation functions and a detailed description of the fitting procedure are provided in the SI. For the $i$-th neuron, the parameters $\alpha_i$, $\beta_i$  and $b_i$ correspond to the slope of the optical response below and above the condensation threshold $P_i^{\text{\rm th}}$, and the background emission intensity, respectively. We optimized the parameters of the neural network using an offline training method (see Methods). 

\subsection{Shape recognition.}
\begin{figure}
     \centering
        \centering  \includegraphics[width=1\textwidth]{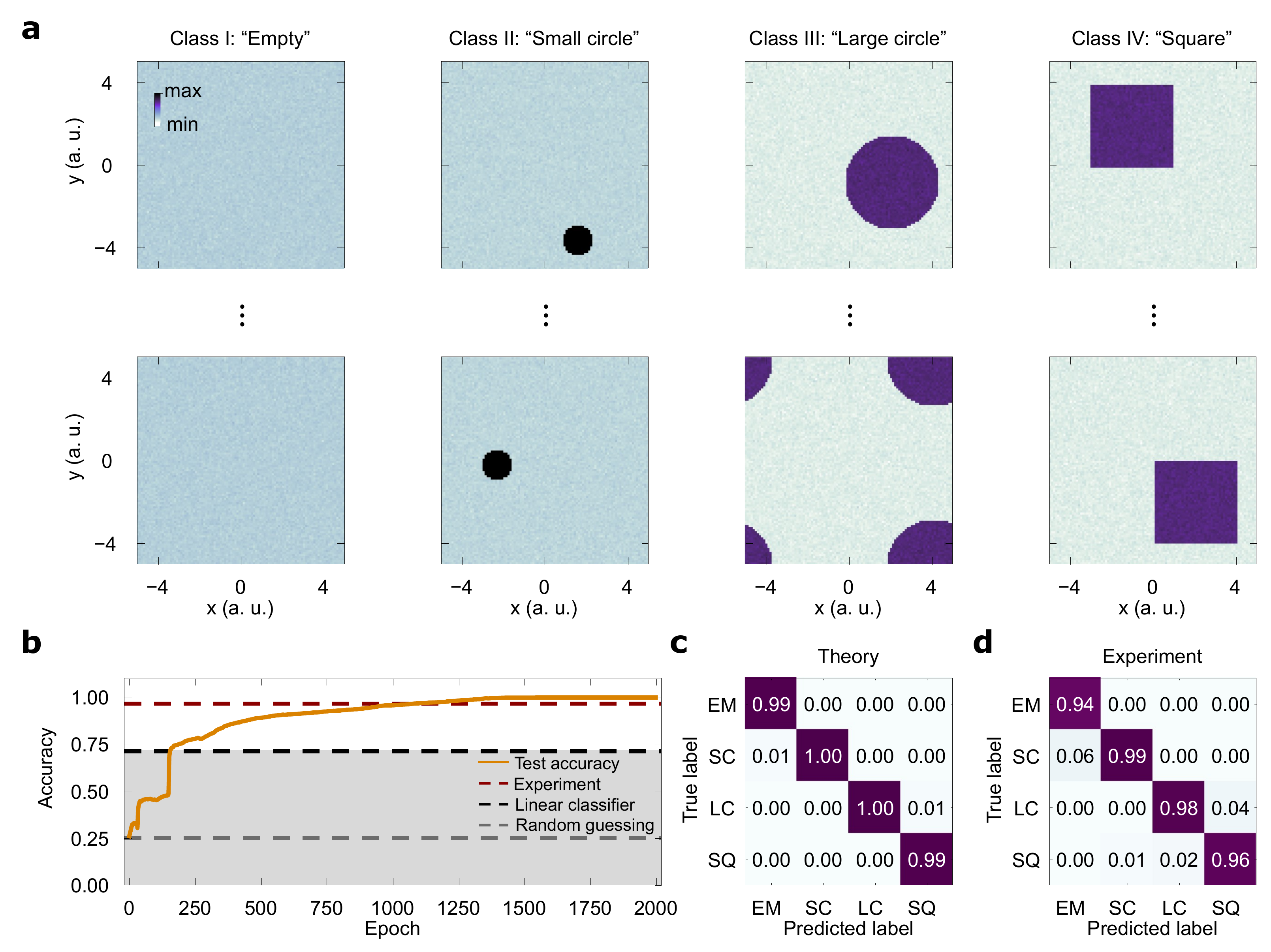}
    \caption{\textbf{ Shape recognition task.}  Panel ~\textbf{a} displays examples of data images used in a testing stage, including empty frames, small circles, squares, and large circles. These objects may appear in different positions within a frame. Panel~\textbf{b}  shows accuracy versus training epoch number for the 4-node network, trained using polariton activation functions. Solid and dashed lines show the training and testing accuracies, respectively. Once the model is trained and tested, activation function from experimental data is used in inference phase with an average accuracy of 96\%, shown by a dark red dashed line. Confusion matrices resulting from the software and experimental inference are respectively shown in panels~\textbf{c} and~\textbf{d}.}
    \label{fig_2}
\end{figure}

As the first application of our room-temperature polariton neural network, we explored the classification of objects based on their shape and size. This task has been chosen specifically to demonstrate the ability of a nonlinear optical network containing just a few nonlinear neurons to tackle a  task of high practical relevance. At the same time, it provides a much higher success rate than a linear network.  We constructed a dataset consisting of four distinct classes of objects: square (SQ), small circle (SC), large circle (LC), and empty sample (EM). These objects were represented as 100 $\times$ 100 pixels images, randomly positioned on a square mesh with periodic boundary conditions. Figure~\ref{fig_2}a provides representative examples of objects from each category. 

According to the offline supervised learning approach, we trained hardware polariton neural networks using a backpropagation algorithm and the fitted polariton neuron responses, see Eq.~(\ref{eq:phi}). Training and testing datasets contained 20,000 and 4,000 samples, respectively, with an equal distribution of classes.  During the training phase, we adjusted the input ($\mathbf{W}_\text{in}$) and output ($\mathbf{W}_\text{out}$) weights of the neural network connecting input and output layers to the optical hidden layer (see Methods). Panel Figure~\ref{fig_2}b illustrates the example of the validation accuracy in the function of the number of training epochs, where the accuracy was determined on the testing dataset. The loss function demonstrates a smooth convergence, while the classification accuracy increased with the number of epochs, confirming good convergence of the training process. We performed statistical characterization by analyzing 100 independent training processes, to confirm the reproducibility of results. Each realization was initialized with different conditions and with varying shuffling seeds for the training dataset. The averaged training and testing accuracies were 94.8$\%$ and 89.7$\%$, with a maximum of 100$\%$ in both cases.

Then we conducted software-supported inference using the experimental data. In this stage, we replaced the neuron output calculated using the analytic activation function used during training with the measured integrated optical signal from the perovskite waveguides. This stage provides a more accurate validation of the hardware polariton neural device in the experimental setting. Panels (c) and (d) from Figure~\ref{fig_2} show the confusion matrix, obtained for the software network, involving an analytical form of polariton activation function, and the realistic case, where experimental data was used. The gap between the accuracy achieved in theory and during the best experimental validation, in which the neural network achieved 96$\%$ of classification accuracy, was equal to 3$\%$. These results demonstrate that a network containing a small number of polariton-based neurons can be used to successfully realize a machine learning task that is challenging for the simple linear classifier (average accuracy at the level of 71$\%$). More details on the dataset, training procedures, and inference methods are provided in the Methods section.
 
\subsection{Binary classification.}
\begin{figure}
     \centering
        \centering  \includegraphics[width=0.90\textwidth]{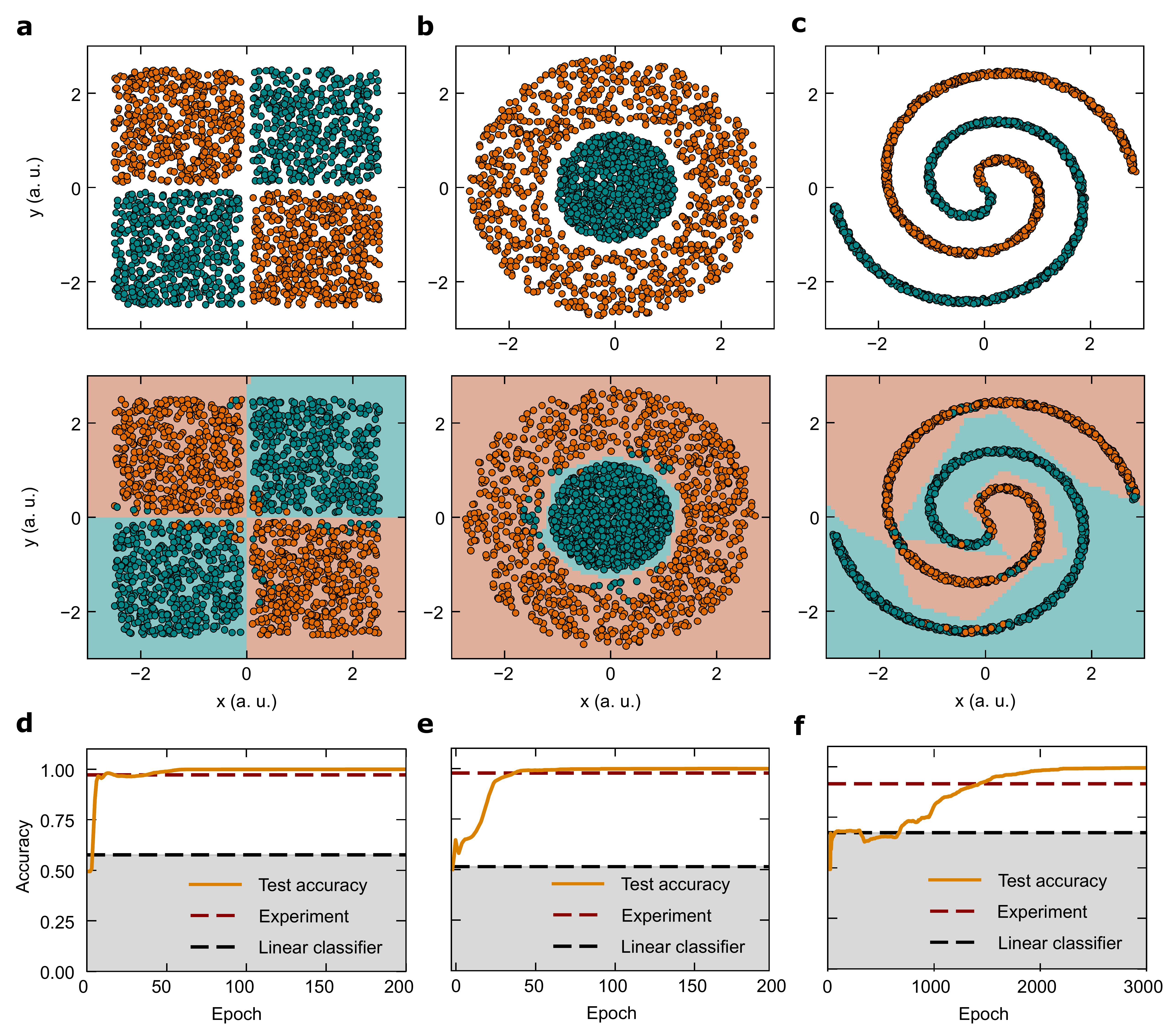}
    \caption{\textbf{ Binary classification task.} Top row panels ~\textbf{a}, ~\textbf{b}, and ~\textbf{c} illustrate the 2,000 training samples used in the training process during the binary classification task. Each point represents a sample within the dataset. Bottom row panels show the results of the experimental inference, where 2,000 testing samples were classified. The points from the testing dataset are displayed together with the decision boundary, represented by a colour map, where two colours represent the class predicted by “ideal” neurons (described by analytical functions). The achieved inference accuracies for datasets represented on panels ~\textbf{a}, ~\textbf{b}, and ~\textbf{c} are 97.25$\%$, 97.75$\%$, and 91.55$\%$, respectively (red dashed lines on panels {\bf{d}}-{\bf{f}}). The slight difference between neural network predictions and experimental inference is a consequence of the “imperfect” hardware implementation of real neurons, which are sensitive to changes in the external environment. Panels {\bf{d}}, {\bf{e}}, and {\bf{f}} illustrate the accuracy across training epochs for datasets a, b, and c, respectively. In each case, the accuracy of the polariton neural network exceeds that of the linear classifier, which is represented by a black dashed line in panels {\bf{d}}, {\bf{e}}, and {\bf{f}}.}
    \label{fig_3}
\end{figure}

Moreover, we validated our perovskite neural network using a binary classification task designed to assess its ability to solve linearly inseparable problems. Such tasks typically require nonlinear transformations of the input data for accurate predictions. To evaluate this, we employed three distinct datasets: rings (dataset a), exclusive OR (dataset b), and spirals (dataset c)~\cite{Smilkov_2017}, as illustrated in Fig.~\ref{fig_3}. In these datasets, the two colors of the points  indicate the two classes to be classified, with each dataset presenting a different level of complexity. For dataset a, the neural network had to learn a radial transformation of the input space to separate the classes. In dataset b, the classes are arranged diagonally, requiring the network to adjust its weights to handle non-linearly separable data. Dataset c posed the greatest challenge, with the classes forming intertwined spirals, demanding more advanced nonlinear transformations and a larger number of neurons to define the intricate decision boundaries required for accurate classification. The classification complexity of each dataset directly influenced the number of neurons required. The polariton neural network achieved good classification with four neurons for the rings and the XOR dataset, and eight neurons for the more complex spirals dataset. This progression aligns with the increasing spatial complexity of each dataset’s feature space. Figure~\ref{fig_3} (lower panels) displays the decision boundaries generated by the network, highlighting its adaptability to each dataset's unique structure. Inference using experimental activations in the hidden layer demonstrated high classification accuracy, achieving 97.25$\%$, 97.75$\%$, and 91.55$\%$,  for datasets a, b, and c, respectively, outperforms linear classifiers, during the training process as was presented in Fig.~\ref{fig_3}d-e.

\section*{Discussion} This work presents the first electro-optic polariton-based neuromorphic system capable of operating at room temperature, where hidden layers of neural networks are implemented using perovskite waveguides. Previously, polariton neural networks were restricted to cryogenic temperatures, relying on optical microcavities constructed from group III-V or II-VI inorganic semiconductors. Our approach, in contrast, takes advantage of the unique properties of halide perovskites, such as their high exciton binding energy and naturally high refractive index, which enable the formation of microcavities without the need for additional structuring. Moreover, previous experiments required fabrication of polariton microcavities with quantum wells positioned between two distributed Bragg reflectors, which relies on costly molecular beam epitaxy technology. In our study,  we replace these expensive, state-of-the-art structures with simple perovskite crystals, fabricated with low-cost and scalable synthesis method~\cite{Kedziora2024}. This approach significantly reduces the fabrication complexity, making it more accessible for practical applications. 

Moreover, perovskite materials facilitate the natural formation of polaritons characterized by strong nonlinearity at a phase transition -- a hallmark of non-equilibrium Bose-Einstein condensation. This property opens up new possibilities for advanced applications in neuromorphic devices, offering improved performance and broader functionality. Additionally, using strongly interacting polaritons in perovskite waveguides eliminates the need for costly optoelectronic data conversion~\cite{Matuszewski_2024}. The optical energy per operation in the inference stage can be estimated from the averaged laser power (7 $\mu$W), the repetition rate of our pulse laser (40 kHz), and the number of required linear and nonlinear operations per neuron ($100^2 +4$ operations in the case of the shape classification task). The resulting optical energy is 17.5 fJ per operation, two orders of magnitude lower than in the case of a binarized neural network based on CdTe microcavities~\cite{Mirek_2021} and comparable to the state-of-the art memristor-based electronic neural networks~\cite{FusionMemristor} and latest optoelectronic systems~\cite{Englund_VCSEL,Taichi}. The perovskite-based system also does not require cooling to cryogenic temperatures, which would contribute to the total energy cost. These estimates do not include, however, the energy cost of electronic circuitry and laser wall-plug efficiency, which need to be taken into account when estimating the total energy efficiency of a large-scale system~\cite{Matuszewski_2024}.

Remarkably, even with only a few polariton neurons, we achieved high accuracy in classifying simple shapes. This may be regarded as a first step in paving the way to lab-on-a-chip ultrafast optical devices, sensors, and energy-efficient computing accelerators. Additionally, recent advances in the design and fabrication of arbitrarily shaped perovskite crystals provide promising prospects for on-chip polariton neuromorphic circuits integrated with laser diodes and $\mu$-lasers. While there is potential for further optimization and development of polariton perovskite circuits to enable scalable, universal optical brain-inspired computing systems, considerable challenges persist. These include material optimization, component integration, and the training of physical neural networks. Overcoming these challenges can open new perspectives for polariton neuromorphic computing applications.
 
In conclusion, we have shown that the unique combination of light and matter properties in room temperature perovskite polaritons makes them an excellent choice for neuromorphic computing. Our work opens new perspectives for the application of perovskite crystals in ultrafast analogue optical neural networks.
\clearpage{}
\newpage{}
\section*{Methods}
\subsection{Perovskite crystal synthesis}
During the the template-
assisted crystallization method, a polydimethylsiloxane (PDMS) template featuring channels of variable width was applied to a commercial glass substrate (1.1~mm thickness). Then a 0.2M solution of CsPbBr$_3$ in dimethyl sulfoxide was distributed on the template. Crystallization occurred in an atmosphere saturated with solvent vapor and at reduced temperatures, ensuring the high quality crystals. For detailed instructions on the preparation of such samples, refer to our previous work~\cite{Kedziora2024} where one can find also material's data including XRD, EDX and absorption measurements. The resulting microwires were utilized in the experiment without any further modification, functionalization, or surface protection. In Figure~\ref{fig_0}b, we present scanning electron microscope (SEM) image of a typical perovskite CsPbBr$_3$ microwire. In addition, we present SEM and confocal fluorescence microscopy images of the waveguides in supplementary information (SI). High quality and reproducibility of microwires support the possible scalability of neuromorphic systems based on monocrystalline perovskites.
\subsection{Experimental setup.}
Measurements used to perform machine learning tasks were conducted at room temperature on a sample consisting of perovskite microwires grown on a glass substrate, without microcavity encapsulation or any protecting layer.  The sample were excited by the picosecond pulsed laserwith a duration $\tau = 1$~ps, 40~kHz repetition rate and a wavelength $\lambda = 435$~nm (2.850~eV). The laser beam power was modulated using the 8-bit SLM in the amplitude mode. The SLM acted as a grayscale filter to vary excitation power in the range of 0.2--14$ \upmu$W with 122 power levels within this range. Laser reflected from the SLM matrix was focused on the sample using a 20$\times$ microscope objective with a numerical aperture of $\mbox{NA} = 0.4$, creating a laser spot diameter of approx. 6~$\upmu$m at $1/e^2$. Next, the emission from the microwire was collected on the 16-bit CMOS camera to measure time-averaged intensity from the edge of the wire. The condensation threshold at approx. 7$\upmu$W of average input beam power corresponds to the average energy consumption of 175~pJ per single neuron. More than 65,000 measurement samples were collected for each condensation site. Plots of the intensity versus excitation power characteristics are available in Fig.~\ref{fig_1} and a detailed scheme of the experimental setup is available in Fig.~S1 in the SI.

\subsection{Nonlinear optical polariton response.}
The form of equation~(\ref{eq:phi}) can be understood based on the coupled kinetic Boltzmann equation, modeling polariton condensation~\cite{Opala_18}. Upon excitation by a short, non-resonant laser pulse $P(t)$, a large population of high-energy excitons and free carriers (electron-hole plasma) is generated. These carriers undergo relaxation, forming a high-energy excitonic reservoir $n_R$ that feeds the polariton condensate $n_C$ through stimulated scattering. This process can be described by the following equations \begin{equation}\label{eq:nC} \frac{d}{dt}n_C=(Rn_R(t)-\gamma_C)n_C(t), \end{equation} \begin{equation}\label{eq:nR} \frac{d}{dt}n_R=P(t)-\gamma_Rn_R(t) -Rn_R(t)n_C(t), \end{equation} where $R$ is the reservoir-condensate scattering rate, and the parameters $\gamma_R$ and $\gamma_C$ represent the decay rates of the exciton reservoirs and polaritons, respectively. An analytical expression for the average polariton density as a function of pump power is obtainable only in the steady-state or adiabatic limit. In such cases, solutions of the above equations reveal two distinct regimes below and above the condensation threshold, corresponding to the absence or presence of a condensed polariton fraction, respectively.
In the absence of a condensate, emission from the system, is linearly proportional to pulse power and inversely dependent on the reservoir decay rate $n_R=\frac{P}{\gamma_R}$.
At a critical pump power $P_{th}=\frac{\gamma_C\gamma_R}{R}$, where system gain surpasses dissipation (the threshold pump intensity), a condensate density builds up. This process is accompanied by a depletion of the reservoir, induced by the presence of stimulated scattering, which in this case takes a constant value of $n_R = \frac{\gamma_C}{R}$. In this regime, the polariton density increases linearly with excitation power $n_C=\frac{P}{\gamma_C}-\frac{\gamma_R}{R}$. Therefore, the integrated emission from the system, given by $I(P)=\alpha n_R+\beta n_C$, where $\alpha$ and $\beta$ are phenomenological parameters that rescale the carrier density intensity to match the optical signal intensity measured in experiments, can be described as
\begin{equation} \label{eq:I}
    I(P) = 
\begin{cases} 
 \frac{\alpha}{\gamma_R}P& \text{if } P < P^{\rm th}_i \\
\frac{\beta }{\gamma_C}P-\frac{\beta \gamma_R}{R} +\frac{\alpha \gamma_C}{R}& \text{if } P > P^{\rm th}_i
\end{cases},
\end{equation}
For time-dependent pulse excitation, the above system of partial differential equations cannot be solved analytically and requires numerical methods, but the qualitative interpretation of the condensation process is the same.

\subsection{Datasets for the shape recognition.} 
The dataset used for the shape recognition task was prepared to evaluate the ability of neuromorphic systems to accurately classify objects with varying shapes, placed at random positions. The rationale behind choosing this dataset comes from the observation that translation-invariant datasets are difficult to process with linear networks. In the case of translational (or permutation) invariance, it can be proven that the only feature that a linear network can use to classify objects is the average input pixel intensity in the image. This is a result of a full equivalence of all input pixels in an invariant dataset. In result, linear network accuracy is on the level of a random guess if all classes have the same average input pixel intensity. To increase the level of difficulty, we chose the areas of the squares and large circles to be equal. This assumption allowed us to test whether the neuromorphic system could recognize both the size and shape of the objects. 

In accordance with the rule mentioned in the main text, each sample was normalized to have the same average pixel intensity. After normalization, we introduced variability by adding noise sampled from a normal distribution with a mean of zero and a standard deviation of $\sigma = 0.05$. Numerical analysis showed that a linear classifier achieved an average accuracy of only 25$\%$ (equivalent to random guessing in a four-class problem), when employing a small batch size and 73$\%$ in the case of the large batch containing 3000 examples. This discrepancy shows that there were some possible other correlations in the dataset that the linear classifier was capable of detecting. Such correlations are a consequence of the methodology employed in dataset preparation and the finite size of our training and testing datasets. Additionally, the tests revealed an imbalance in prediction difficulty. Distinguishing between objects from the EM and SC classes was relatively straightforward for a small neural network while differentiating between objects with the same area but different shapes (LC and SQ) was significantly more challenging. As a result, during the training process, we observed that the system only began recognizing object shapes (squares and circles of the same size) when neural network accuracy exceeded 75$\%$. 

\subsection{Offline training and inference stage.} 
This process was entirely software-based, utilizing Python packages such as TensorFlow to implement the backpropagation algorithm.  During training, we optimised the input $\mathbf{W}_\text{in}$ and output $\mathbf{W}_\text{out}$ weight matrices by minimising the loss function $L$. This process was based on a supervised learning method performed using the ADAM optimiser.

In the shape recognition task, we set the learning rate to $25\cdot10^{-5}$ and used a batch containing $3\cdot 10^{3}$ examples. In this task, the number of epochs was equal to $2\cdot 10^{3}$. Additionally, we employed an $L_2$ regularization method with $10^{-5}$ scaling factor. We truncated the values of the weights to the range from $-0.05$ to $0.05$ to ensure a symmetric distribution of weights in the training process. To achieve optimal classification performance we employed a focal loss function defined as follows
\[
L(p_t)=-\alpha_t(1-p_t)^\gamma\log{(p_t)},
\]
where $\gamma$ is a focusing parameter,  $\alpha_t$ defines the weighting factor and $p_t$ is a probability of ground truth class prediction. In our case $\alpha_t =0.5$ and $\gamma_t =3$. 

 The neural network output $y$ for a given input vector $\mathbf{x}$ was described by
\[
y(\mathbf{x}) = f(\mathbf{W}_\text{out} \phi(\mathbf{W}_\text{in} \mathbf{x})),
\]
where $\phi(\cdot)$ and $f(\cdot)$ are activation functions of hidden and output layers, respectively and $\mathbf{b}$ defines a bias vector. During training, we move the threshold of the activation function to the centre of the coordinate system in the neuron input-output space.  For the $k$-th input example, denoted by the input vector $\mathbf{x}^{(k)}$, the optical input intensities acting on the neurons are given by $\mathbf{I}^{(k)} = \mathbf{W}_\text{in} \mathbf{x}^{(k)}$. As illustrated in Figure 1a, the activation functions of the polariton artificial neurons are slightly different. Therefore, the activations over the hidden layer are defined by \[\phi(\mathbf{I}) = [\varphi_1(I_{1}), \varphi_2(I_{2}), \varphi_3(I_{3}), \varphi_4(I_{4})].\] 
In binary classification, we introduce an additional bias term, denoted as $\mathbf{b}$, in the output layer to improve neural network accuracy. Instead of using the focal loss function, we employ a standard mean square error (MSE) loss between the true labels $y^{(k)}$ and the predictions $\hat{y}^{(k)}$, described by the following equation
\[
L=\frac{1}{n}\sum_{k=1}^{n}({y^{(k)}-\hat{y}^{(k)}})^2,
\]
where $n$ is the number of samples. During the training phase, the following sets of parameters were employed: $S_{(a)}=\{1000,0.025,200\}$, $S_{(b)}=\{1000,0.025,200\}$, and $S_{(c)}=\{500,0.001,3000\}$, which defined the batch size, learning rate, and number of training epochs, respectively, for datasets a, b, and c. To classify the spiral dataset, we employed a neuron multiplexing method, utilizing eight neurons. The activation function of the hidden layer was defined as \[\phi(\mathbf{I}) = [\varphi_1(I_{1}), \dots, \varphi_4(I_{4}),\varphi_1(I_{5}), \dots, \varphi_4(I_{8})].\]
It should be noted that the offline training method yields excellent results for a single hidden layer feed-forward neural network based on isolated polariton neurons~\cite{Opala_2022}. Nevertheless, it should be noted that for more advanced configurations mimicking a recurrent neural network, where neurons within a hidden layer are interconnected, training can be more time-consuming and may require advanced techniques, such as surrogate models, physics-aware training~\cite{Wright_2022}, equilibrium propagation~\cite{Scellier_2017} or physically informed neural networks~\cite{Karniadakis_2021}. In the inference stage, we used the true experimental measurements as the activation function in the hidden layer, while the linear part of the network was implemented in the software.
 
In the inference phase, we used data collected during the characterization of the polariton neurons. First, the activations of the neurons for inputs $\mathbf{W}_{\text{in}} \mathbf{x}$ within the experimental range were replaced by the measured polariton responses corresponding to the nearest optical input intensity recorded in the dataset. During this procedure, we used 4$\times$936 measured points, corresponding to 12 power-intensity scans. Each point could be selected multiple times. This approach allowed us to identify which machine learning tasks the optical system could effectively solve without repeating experiments for each training and testing phase. Additionally, it helped determine the optimal scaling and conditions for tuning the network parameters.

\clearpage{}
\newpage{}

\clearpage{}
\newpage{}

\clearpage{}
\newpage{}

\newpage{}
\clearpage{}

\section*{Data availability}
\noindent All data that supports the conclusions of this study are included in the article. The data presented in this study are available from the corresponding author upon reasonable request.

\section*{Acknowledgments} 
\noindent  We would like to thank Timothy C. H. Liew, and Piotr Kapuściński for stimulating discussions.  Additionally, we would like to thank Jacek Szczytko for sharing laboratory space and resources and support during the project. This work was supported by the National Science Center, Poland, under the following projects: 2022/47/ B/ST3/02411,  2021/43/B/ST3/00752, 2023/49/N/ST3/03595, 2023/49/B/ST3/00739, 2024/52/C/ST3/00324 and financed by the European Union EIC Pathfinder Open project “Polariton Neuromorphic Accelerator” (PolArt, Id: 101130304). A.O. acknowledges support from the Foundation for Polish Science (FNP). M.E. and A.Sz. acknowledge the support by the statutory funds of the Łukasiewicz Research Network – Institute of Microelectronics and Photonics. Finally, we would like to thank Mateusz Król for creating the 3D graphics for this manuscript.

\section*{Author contributions}
\noindent A.O., K.T., M.K., M.M. and B.P. conceived the idea, A.O., S.S. and A.R. performed numerical simulation and optimization, K.T. performed the optical experiments, M.E. and A.Sz. made GaAs masters, M.K. grew perovskite crystals, A.O. and M.M. wrote the manuscript with significant input from K.T., M.K., M.F. and B.P.. The project was supervised by A.O., M.M. and B.P..
\section*{Competing interests} 
\noindent The authors declare no competing interests.

\section*{Additional information} 
\noindent{\bf Correspondence and requests for materials} should be addressed to B.P.

\section*{References}
\vspace{10mm}
\bibliographystyle{naturemag}
\bibliography{bib}

\end{spacing}
\newpage{}
\clearpage{}

\title{\bf{Supplementary Information}}

\setlength\parskip{0.25em}

\begin{spacing}{1.125}


\begin{spacing}{0.8}

\tableofcontents
\end{spacing}
\newpage

\section{Microscopic characterization of perovskite waveguides}

\begin{figure}[ht!]
		\centering
		\includegraphics[width=1\linewidth]{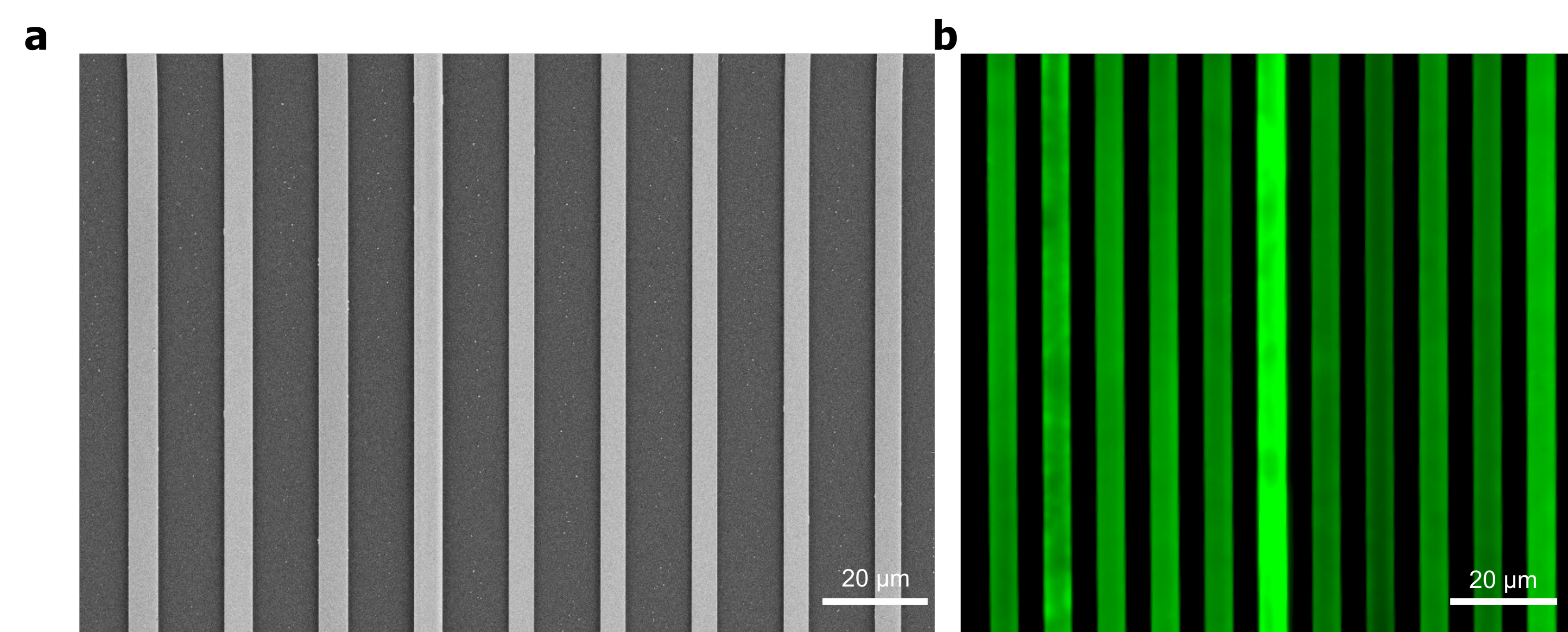}
		\caption{\bf{Microscope images of CsPbBr$_3$ microwires.} a SEM and b scanning fluorescent images.} 
		\label{im:SI-SEM}
	\end{figure}

Fig.~\ref{im:SI-SEM}a shows scanning electron microscope (SEM) images of the fabricated CsPbBr$_3$ microwires. The microwires are arranged in a grid of evenly spaced structures, each of which exhibits excellent quality. This precise arrangement underscores the high potential of this technology for scalable integration into large networks.

The scanning fluorescence microscope images of the wires illustrated in Fig.~\ref{im:SI-SEM}b (arranged in a grid with different spacing than in Fig.~\ref{im:SI-SEM}a) highlight the wires' quality in terms of emission properties, showing high homogeneity and minimal scattering defects. Enhanced emission at the microwire edges marks the crystal facets, where the photonic field leaks from the naturally formed cavity.
 
\section{Detailed scheme of the measurement setup}

\begin{figure}[ht!]
		\centering
		\includegraphics[width=1\linewidth]{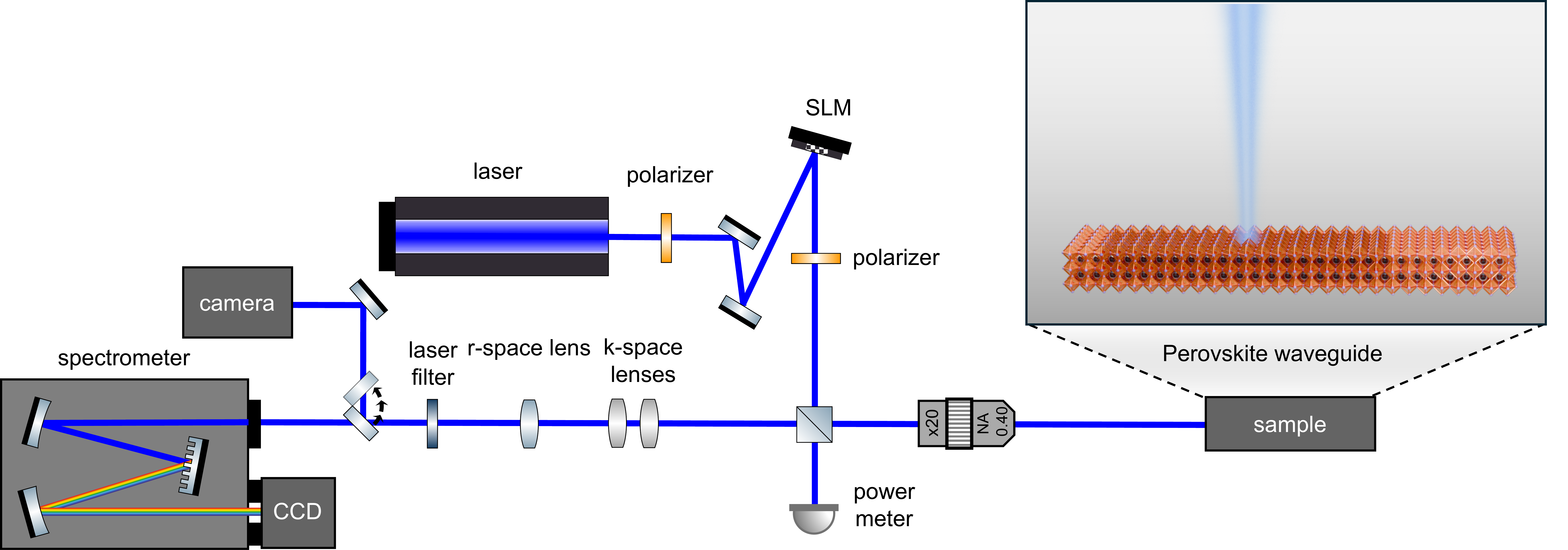}
		\caption{{\bf{Main experimental setup.}} Experimental setup for measuring emission from a perovskite microwire, using SLM.}
		\label{im:setup1}
	\end{figure}

The setup used in the experiment to collect training and testing characteristics of each neuron with ReLU activation function is described in the Methods section in the main text and schematically shown in Fig.~\ref{im:setup1}. In this experimental setup, after collecting the emission dependence from the first excitation region (corresponding to neuron 1), the position of the laser acting on the perovskite wire was changed. This way, the detailed nonlinear emission characteristics from four different locations on the sample were obtained. This was essential for studying the nonlinear neuronal activations and the degradation process of the perovskite crystals. 

\begin{figure}[ht!]
		\centering
		\includegraphics[width=1\linewidth]{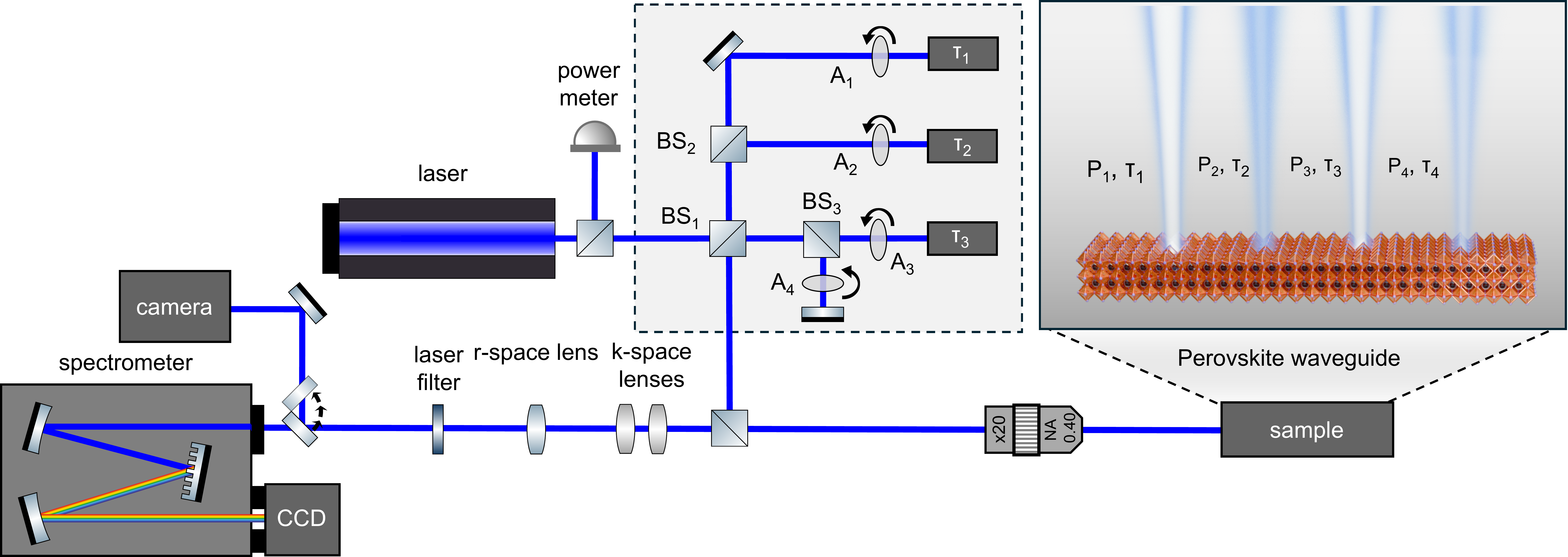}
		\caption{{\bf{Experimental setup for scalable 4-spot excitation.}} Experimental setup for measuring emission of the four spatially and temporally separated polariton condensates created with the perovskite microwire.}
		\label{im:setup2}
	\end{figure}

The setup shown in Fig.~\ref{im:setup2} was used to optically demonstrate the simultaneous creation of four (spatially and temporally independent) polariton condensates acting as network of neurons. To achieve this the laser beam passed through an experimental setup designed to split beam into four parts with tunable power. The setup consisted of 50:50 beam splitters [(BS$_1$, BS$_2$, BS$_3$)] and adjustable gray filters (A$_1$, A$_2$, A$_3$, A$_4$) used to independently adjust the power of every beam. This system incorporated delay lines ($\tau_1$, $\tau_2$, $\tau_3$) that were engineered to separate four laser pulses in space and time. By adjusting the movable mirrors on each delay line, the temporal delay between individual pulses was tuned. The four spots were focused on the sample with 20$\times$ microscope objective with a numerical aperture of $\mbox{NA} = 0.4$ (Nikon TU Plan ELWD) and signal was collected on sCMOS camera (Andor Marana).

\clearpage{}

\section{Polariton neuron activation functions, and their stability}

To fit experimental data with an analytical form of polariton activation function we use least-squares algorithms implemented in the SciPy Python package. For each from measured condensation sites, we obtain polariton neuron threshold $P_{th}=\{7.34,6.36,5.24,7.18\}$ given in ~$\upmu$W, the linear slope below activation $\alpha_i=\{0.02,0.01,0.03,0.01\}$, linear slope above activation $\beta_i=\{0.15,0.14,0.29,0.35\}$, and background intensity $b_i=\{0.21, 0.16, 0.24, 0.14\}$.

\begin{figure}[bht!]
		\centering
		\includegraphics[width=0.8\linewidth]{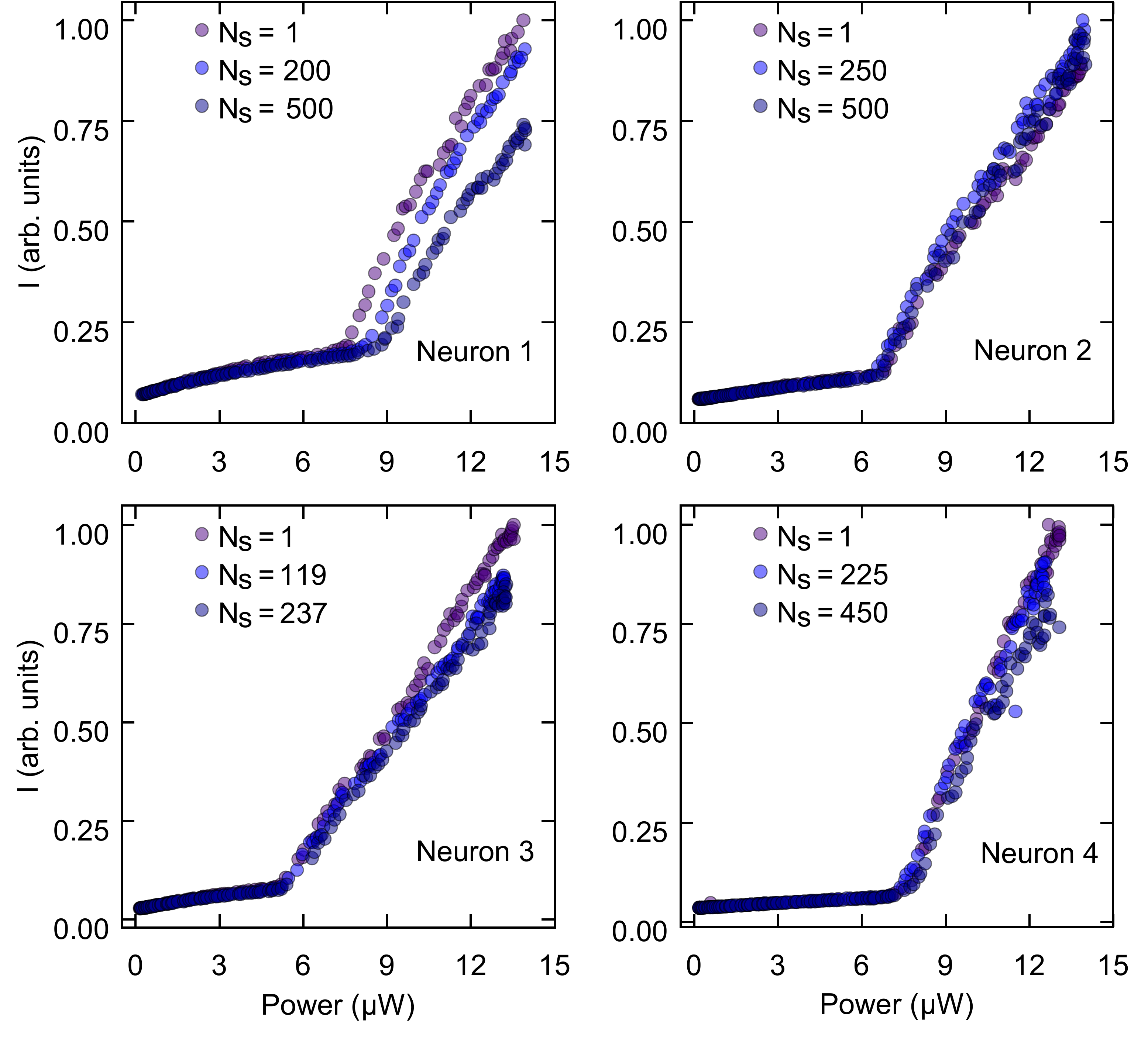}
		\caption{{\bf{Perovskite crystal degradation.}} Material degradation observed in sequential input-output power measurements for each neuron. Parameter $N_s$ corresponds to the scan number. }
		\label{im:degradation}
\end{figure}
 
During our optical response measurements, a gradual degradation process in the perovskite material was tested. To capture this effect, we conducted multiple scans of each neuron's nonlinear response to varying excitation power, with over 500 sequential photoluminescence measurements taken at each position. This process spanned over more than an hour, allowing us to observe subtle changes in the material properties over time. Fig.~\ref{im:degradation} shows how each neuron's activation function evolves throughout the experiment. The violet, blue and dark blue points correspond to the initial, intermediate and final scans, respectively. Notably, for neurons 2, 3, and 4 the activation function remains fairly stable across scans; however, we observe a steady change in the slope of the nonlinear response and threshold, suggesting an impact of the degradation on the dynamic response of the condensate. In contrast to neurons 2, 3, and 4 neuron 1 exhibited the most rapid degradation process.  

\begin{figure}[ht!]
		\centering
		\includegraphics[width=0.8\linewidth]{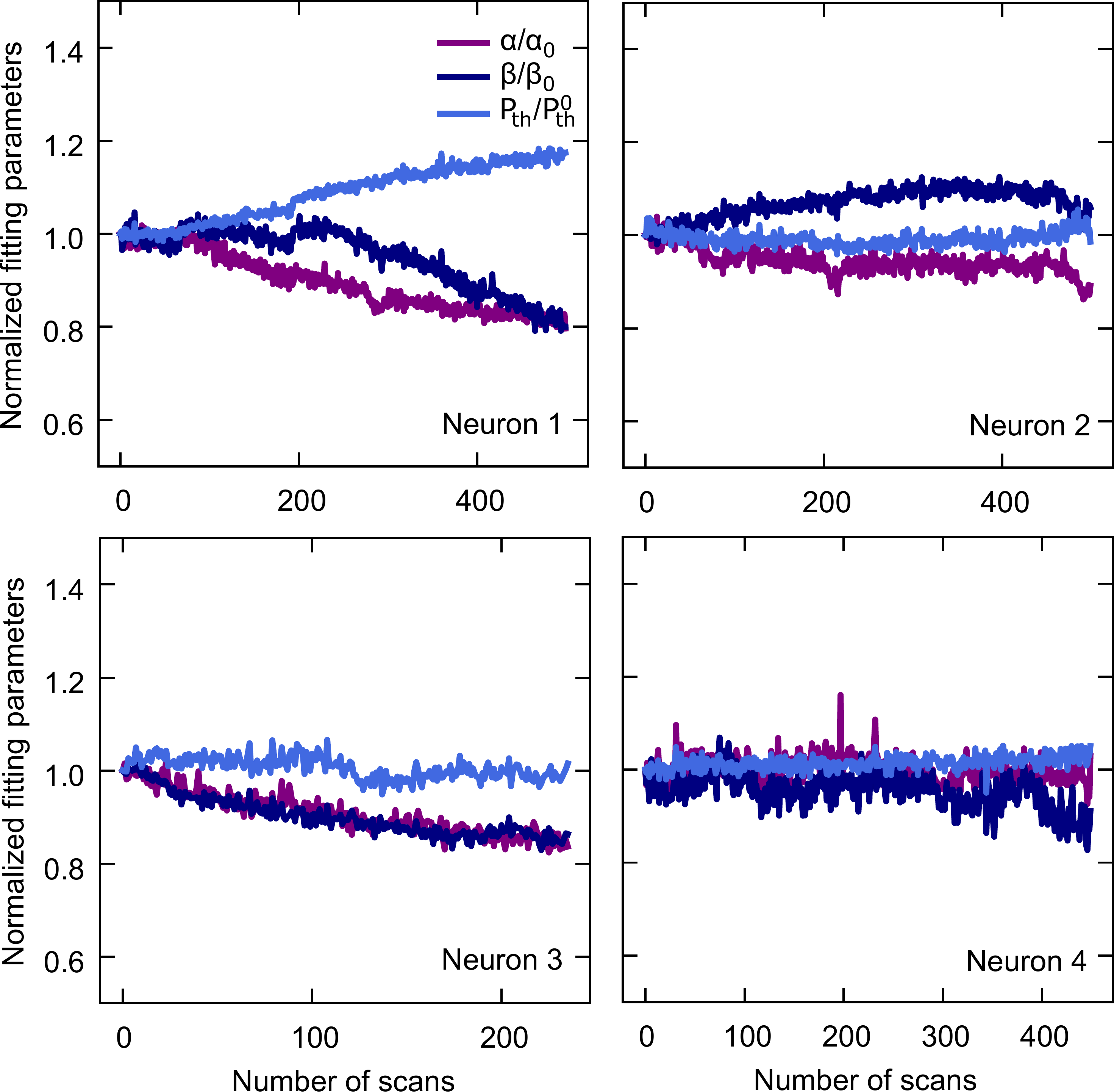}
		\caption{{\bf{Variations of polariton neurons activation functions observed during the scanning process.}} Normalized slope and neuron activation threshold variations in the input-output characteristics of each neuron during successive power scans.The parameters $\alpha_0$, $\beta_0$ and $P_{th}^{0}$ correspond to the parameters obtained from the first scan.}
		\label{im:degradation-gradient}
	\end{figure}

To further analyze this effect, we calculated the slopes and threshold changes across the input-output scans, which are shown in Fig.~\ref{im:degradation-gradient}. The high number of iterations was crucial for accurately characterising nonlinear neural activations and monitoring perovskite degradation under prolonged exposure, highlighting material stability's impact on the robustness of perovskite-based neural networks.

\end{spacing}

\end{document}